\documentclass[useAMS,usenatbib]{mn2e}
\usepackage{amsmath,amssymb}
 \usepackage{graphicx,color}
 \usepackage{bm}
\newcommand{\be}{\ba}    % for lazy typers
\newcommand{\ee}{\ea}
\newcommand{\ba}{\begin{eqnarray}}
\newcommand{\ea}{\end{eqnarray}}
\newcommand{\nn}{\nonumber \\}

\def\E{\mathcal E}
\def\R{\mathcal R}
\let\e=\varepsilon
\let\d=\delta

\title[Resonances and bifurcations in systems with elliptical equipotentials]
{Resonances and bifurcations in  systems with elliptical equipotentials}
\author[Antonella Marchesiello and Giuseppe Pucacco]
{Antonella Marchesiello$^{1}$\thanks{E-mail: anto.marchesiello@sbai.uniroma1.it} and 
Giuseppe Pucacco$^{2,3}$\thanks{E-mail:
pucacco@roma2.infn.it}\\
$^{1}$Dipartimento di Scienze di Base e Applicate per l'Ingegneria -- Universit\`a di Roma ``la Sapienza"\\
$^{2}$Physics Department, University of Rome ``Tor Vergata", I00133 Rome\\
$^{3}$INFN, Sezione Roma Tor Vergata
}
\begin{document}

\date{Received 19/09/2012. Accepted 08/10/2012. }

\pagerange{\pageref{firstpage}--\pageref{lastpage}} \pubyear{2012}

\maketitle

\label{firstpage}

\begin{abstract}

We present a general analysis of the orbit structure of 2D potentials with self-similar elliptical equipotentials by applying the method of Lie transform normalization. We study the most relevant resonances and related bifurcations. We find that the 1:1 resonance is associated only to the appearance of the loops and leads to the destabilization of either one or the other normal modes, depending on the ellipticity of equipotentials. Inclined orbits are never present and may appear only when the equipotentials are heavily deformed. The 1:2 resonance determines the appearance of bananas and anti-banana orbits: the first family is stable and always appears at a lower energy than the second, which is unstable. The bifurcation sequence also produces the variations in the stability character of the major axis orbit and is modified only by very large deformations of the equipotentials. Higher-order resonances appear at intermediate or higher energies and can be described with good accuracy.

\end{abstract}


\section{Introduction}

In investigating the orbit structure of a galactic potential, we are often interested in some particular feature of its general layout. We may mention: the birth and/or disappearance of specific orbit families; their stability nature; the phase-space fraction occupied by invariant tori around stable periodic orbits, etc. \citep{bt}. In integrable systems these features are uniquely determined by the integrals of motion \citep{dzb}: only a limited number of orbit families exist and their possible bifurcations occur at isolated critical values of the conserved functions giving the integrals. 

On the other hand, the dynamics of generic systems are not integrable. There are several bifurcations with a proliferation of periodic orbit families and sooner or later a transition to a stochastic behavior. Stochasticity, if not limited to small regions of phase space, leads to chaos \citep{conto2}. However non-integrable dynamics do not prevent regular behavior: significant parts of phase space can be layered with invariant surfaces and in many instances a generic system as a whole can be quite similar to an integrable system \citep{hh}. In these circumstances perturbation approaches can be devised to describe the features of the system \citep{gu}.

A powerful perturbation method is that based on  {\it Hamiltonian Normal Forms} \citep{DB}. Typically, the application of this method is based on three steps: 

\noindent
1. Construction of a new Hamiltonian (the `normal form') by means of a canonical transformation suitable to capture a peculiarity of the system under study.

\noindent
2. Use of the normal form to investigate in the simplest way the particular feature of the system we are interested in.

\noindent
3. Inversion of the transformation to describe this feature in terms of the original parameter of the system allowing, if possible, the comparison with observational data.

In some cases, steps 2. and 3. can be reversed but, usually, working with the normal form in the normalization coordinates is easier and/or more effective. 

We recall that coping with non-integrable dynamics through perturbation theory often means to try to compute non-existing quantities. To clarify this seemingly absurd statement, we can say that in any perturbative approach dynamical quantities are expressed as series in some (small) parameter. Physical properties of the original system (e.g. the gravitational potential) give convergent series in a suitable neighborhood. The normalization procedure provides expansion series for quantities approximating phase-space invariants. As a matter of fact, these series do not converge: if they would, this should imply that {\it global} invariants existed and that the dynamics should be everywhere regular, which is not. However, due to the {\it asymptotic} nature of these series \citep{SV}, their truncations can be used to approximate {\it local} invariant surfaces responsible for regularity. The change in their geometric properties caused by the variation of physical parameters allows us to locate bifurcations of new orbit families. A remarkable fact is that often the neighborhood of asymptotic semi-convergence is large enough to provide reliable information in a wide range of physical parameters and several features of non-integrable dynamics can be accurately predicted \citep{pbb}. Technical issues do not always lead to a  straightforward application of the method: just to mention a few, we recall that among the many features characterizing the real system, the normal form is able to describe only a limited subset of them, typically in the neighborhood of a given resonance \citep{bbp}. Another issue is that there is no definite strategy to predict the best truncation of the series expansions; moreover, it can be not easy to re-express them in terms of observables. Generic cases (colloquially, models with `many' parameters) are cumbersome: mathematicians have therefore introduced simplifying techniques (singularity theory, catastrophe theory, etc.). However, in spite of their power and elegance, they are even more difficult to use so that, in applications, they are still not so useful. We prefer to stay on `standard' methods. 

Aim of the present paper is to offer a well defined setting in which many of the technical issues listed above are addressed and solved. We will see how to construct normal forms for the dominating resonances, how to use higher-order expansions to predict bifurcation thresholds and stability transitions and will show circumstances in which these quantities can be computed in a `large' range of parameters. Resonance between two non-linear oscillations is the source of non-trivial dynamics \citep{conto1,CM,fv,Bi}. \citet{zm} made a general analysis of the symmetric 1:1 resonance with the averaging method. The method of normal form is more flexible in treating generic resonances requiring higher-order computations and, when applied to the same models, the results are identical to those of the averaging method \citep{MP}. However, when a comparison with numerical results (see e.g. \citet{mes}) requires precise predictions, higher-order computations are necessary \citep{bbp}. The setting in which we work is that of potentials with similar concentric ellipsoidal equipotentials. To shed light on the methods and to limit the algebraic complications we limit the treatment to 2-DOF non-rotating systems. Explicit formulas for the bifurcation thresholds of the main periodic orbits are computed in terms of the energy for a family of models with two shape parameters. Additional ellipsoidal symmetry-breaking perturbations are included. We also discuss the relation with other issues like St\"ackel fits to separable systems, surfaces of section, order and chaos, etc. In particular, we stress how the asymptotic nature of the method allows us to make reliable predictions in a domain much larger than expected on the basis of standard perturbation arguments.

The plane of the paper is as follows: in Section 2 we introduce the procedure to construct the approximating integrable systems by recalling the method of the {\it Lie transform} \citep{gs,gior}; in Sections 3 and 4 we apply this approach to investigate the main aspects of the dynamics in a symmetry plane of a triaxial ellipsoid \citep{bbp}, obtaining first-order estimates of the bifurcation thresholds of the 1:1 and 1:2 periodic orbits; in section 5 we analyze higher-order cases; in Section 6 we discuss further developments and hints for other applications and in Section 7 we conclude.

\section{Resonant Hamiltonian normal forms}

Let us consider a two degree of freedom system with a smooth potential with an absolute minimum in the origin, symmetric under reflection with respect to both  coordinate axes. The Hamiltonian is given by
\ba \label{Hamiltonian}
	\mathcal{H}(\bm w)=\frac12(p^2_x+p_y^2) + \mathcal{V}^{(N)}(x^2,y^2)
\ea 
where with $\bm w$ we collectively denote the phase-space variables and we assume that the potential can be expanded as a truncated power series
\ba\label{seriesp}
 \mathcal{V}^{(N)}(x^2,y^2)\equiv \sum_{k=0}^N  \mathcal{V}_{2k}(x^2,y^2)
\ea
where
\ba\label{potential}
 \mathcal{V}_{2k}(x^2,y^2)=\sum_{j=0}^{k+1} C_{2j,2(k-j)}x^{2j} y ^{2(k-j)}.
\ea
The truncation order $N$ and the coefficients $C_{ij}$ are determined by the problem under study.

In particular, we are interested in a fairly general class of potentials with self-similar elliptical equipotentials of the form
\ba
\label{pota}
\mathcal{V} (x,y;q,\alpha) =\left\{
                    \begin{array}{ll}
                      \frac{1}{\alpha}\left(1 + x^2 + \frac{y^2}{q^2}\right)^{\alpha/2},\;\; & 0<\alpha<2 \\
                      \frac12 \log\left(1 + x^2 + \frac{y^2}{q^2}\right),\;\; & \alpha = 0.
                    \end{array}
                  \right.
\ea
The ellipticity of the equipotentials is determined by the parameter $q$: for short, we will speak of an `oblate' figure when $q < 1$ and a `prolate' figure when $q > 1$. The profile parameter $ \alpha $ determines the behavior at large radius. 

The family of potentials  (\ref{pota}) can be expanded in a series of the form (\ref{potential}), or more simply of the form
\ba\label{seriese}
 \mathcal{V}^{(N)}(x^2,y^2;q)\equiv \sum_{k=0}^N  B_{k} s^{2(k+1)}(q)
\ea
where we have introduced the `elliptical radius'  
\ba
s(q) = \sqrt{x^2+ \frac{y^2}{q^2}}. \ea
With unit `core radius' we can put $B_0 = 1/2$ and, for the class (\ref{pota}), the first two coefficients of the higher-order terms are
\ba
B_1=-\frac{2-\alpha}{8},\quad
B_2=\frac{(2-\alpha)(4-\alpha)}{48}.\ea
Another interesting case is that of the `flattened isochrone' \citep{EZL}:
\ba
B_1=-\frac14,\quad
B_2=\frac{5}{32}.\ea
Each term in the series is given by an even power of the basic elliptical radius and the Hamiltonian (\ref{Hamiltonian}) can be treated in a perturbative way as a non-linear oscillator system. 

To find the normal form we have first of all to `prepare' the Hamiltonian. We start by introducing a small parameter $\e>0$ and, by performing a `blowing-up' of the phase-space by means of the transformation
\ba
\bm{ w}\rightarrow\e^{-1}\bm{ w},
\ea
we rescale the Hamiltonian \eqref{Hamiltonian} according to
\ba
\widetilde H(\bm w) \doteq \e^{-2} \mathcal{H}(\bm w) =\frac12(p_x^2+p_y^2)+\sum_{k=0}^N\e^{2k}\mathcal{V}_{2k}(x^2,y^2).
\ea
With this trick we assign an order to the terms in each series without making an explicit reference to the extent of the neighborhood of the equilibrium. After a further scaling 
\ba p_{x} & =  & \sqrt{\omega_1} p_1, \quad  x = x_1/\sqrt{\omega_1}, \label{sca1} \\
      p_{y} & =  & \sqrt{\omega_2} p_2, \quad  y = x_2/\sqrt{\omega_2}, \label{sca2}\ea
      where
      
      \ba \label{uf}
\omega_1 \doteq  \sqrt{2 B_0} = 1,  \quad
\omega_2 \doteq  \sqrt{2 B_0}/q = 1/q, \ea
      the original Hamiltonian system (\ref{Hamiltonian}) is put into the form 
            \ba
      \widetilde H(\bm{ w}) =\sum_{n=0}^{N} \e^{2n} \widetilde H_{2n},\ea
      where we still use $\bm{ w}$ to denote phase-space variables. We then have
      
      \ba\label{Hzero}
\widetilde H_{0} = \frac12 \left(\omega_1 (p_1^2 + x_1^2) + \omega_2 (p_2^2  +x_2^2) \right)
\ea
and $\widetilde H_{2j} (\bm{ w}), j>0,$ are essentially the higher (than the second) order terms of the potential. We are interested in the behavior of the system `around' $m/n$ resonances with $m,n \in {\mathbb N}$: in general our frequency ratio $q$ is an irrational number and the unperturbed system is non-resonant. However the nonlinear higher-order terms produce a passage through resonance with interesting dynamics. To describe this phenomenon we introduce a `detuning' parameter $\d$ \citep{fv,zm} such that the frequency ratio is
written as
\ba \label{det}
\frac {\omega_1}{\omega_2}=q=\frac{m}{n}+\d.
\ea
The detuning parameter is treated as a term of order two in $\e$ ($\d=\tilde\d\e^2$) and considered `small'. After a further rescaling
\ba\label{scah}
H\equiv \frac{n}{\omega_2} \widetilde H= 2 q  \widetilde H
\ea
and noting that, in view of \eqref{det}, we have
\ba\label{dete}
\frac{1}{q}=\frac nm - \frac{n^2}{m^2}\tilde\d \e^2 + \frac{n^3}{m^3}\tilde\d^2 \e^4 + \dots,\ea
by collecting terms up to  order $2N$ in $\e$, we finally put the Hamiltonian into the
form
\ba
\label{Hr}
 H(\bm w)=\sum_{k=0}^{N}\e^{2k}H_{2k}(\bm w)
\ea
where the unperturbed term (in {\it exact} resonance) is given by
\ba\label{Hzmn}
H_0=\frac12 m(p_1^2 + x_1^2)+\frac12 n(p_2^2 + x_2^2).
\ea
The system is now ready for a standard resonant normalization: it undergoes a canonical transformation to new variables $
\bm W (\bm w),$ such that the new Hamiltonian is
\ba\label{HK}
     K(\bm W)=\sum_{n=0}^{N}\e^{2n}K_{2n}={\rm e}^{{\cal L}_G} H(\bm w),\ea
    where the linear differential operator 
\ba\label{diff}
{\rm e}^{{\cal L}_G} = \sum_k \frac1{k!} {\cal L}_G^k,\ea
associated to the generating function $G(\bm w)$, is defined by its action on a generic function $F(\bm w)$ by the Poisson bracket:
\ba
{\cal L}_G F \doteq \{F,G\}.\ea
To construct $K$ starting from $H$ is a recursive procedure exploiting an algorithm based on the Lie transform \citep{DB,gior}. A short account useful for the present purpose is given in \citet{MP}. To proceed we have to make some decision about the structure the new Hamiltonian must have, that is we have to chose a {\it normal} form for it. We construct the new Hamiltonian in such a way that it admit a new integral of motion, that is we consider a certain function, say $I(\bm w)$, and impose that
\ba\label{NFD}
\{K,I\}=0.
\ea
The usual choice (but not necessarily the only one) is that of assuming
\ba\label{UC}
I=H_{0}=K_0
\ea
so that the function (\ref{Hzmn}) plays the double role of fixing the specific form of the transformation and assuming the status of second integral of motion. 

Formally, a more direct way of applying this method is by using smarter coordinates which greatly simplify the procedure. A first choice is that given by the complex coordinates
\ba\label{cc}
z_1=p_1+i x_1,\quad
z_2=p_2+i x_2,
\ea
leading to a normal form $K(z_1,z_2, \bar z_1, \bar z_2)$ so that, for example,
\ba\label{cale}
H_{0}=K_0=\frac12 (m z_1 \bar z_1 + n z_2 \bar z_2).
\ea
A second useful choice is the action-angle--{\it like} variables ${J}_{a}, {\theta}_{a}$ defined through the transformations
\ba
z_{a}= i \sqrt{2J_{a}}e^{-i \theta_{a}}, \;\;\; a=1,2.
\ea
In this way we have
\ba\label{calei}
H_0 =K_0 = m J_1 + n J_2, \ea
so that
\ba\label{LHAA}
 {\cal L}_{H_0} = m \frac{\partial}{\partial \theta_1}  + n \frac{\partial}{\partial \theta_2} . \ea
 With the choice (\ref{UC}), condition (\ref{NFD}) translates in the necessary condition 
\ba\label{LHK}
 {\cal L}_{H_0} K =0\ea
 the new Hamiltonian must satisfy. 
Since a generic polynomial series turns out to be a Fourier series in the angles with coefficients depending on the actions,
the typical structure of the resonant normal form (truncated when the first resonant term appears) is \citep{SV}
\begin{eqnarray}\label{GNF}
K&=&m J_1+n (J_2 + \e^2 \tilde\d J_1) + \sum_{k=1}^{m+n-1} \e^{2 k} {\cal P}^{(k+1)}(J_1,J_2)+\nonumber \\
&&\e^{2(m+n-1)} A_{mn} J_1^{n} J_2^{m} \cos [2(n \theta_{1}- m \theta_{2})], 
\end{eqnarray}
where ${\cal P}^{(j)}$ are homogeneous polynomials of degree $j$ whose coefficients may depend on $\delta$ and the constant $A_{mn}(q,\alpha)$ is the coefficient of the resonant term. It is easy to check that this is the most general form of a phase-space function of degree $m+n$ in the actions which stays in the kernel of  $ {\cal L}_{H_0}$ as given by (\ref{LHAA}).
In these variables, the second integral is given by (\ref{calei}) and the angles appear only in the resonant combination $n \theta_{1}- m \theta_{2}$: for a given resonance, these two statements remain true for arbitrary 
\be
N>N_r \doteq m+n-1,\ee 
where $N_r$ is the order of the resonance. New variables `adapted to the resonance' \citep{SV} are introduced by means of the quasi-canonical transformation,
\ba\label{calr}
\begin{array}{ll}
{\cal E}=&\frac{\lambda}{m^2+n^2} (m J_1 + n J_2), \\
{\cal R}=&\frac{\lambda}{m^2+n^2} (n J_1 - m J_2), \\
\psi=&\mu(n \theta_{1} - m \theta_{2}), \\
\chi=&\mu(m \theta_{1} + n \theta_{2}).
\end{array}
\ea
The transformation is canonical when $\lambda=1/\mu$, but other choices can be  convenient to simplify formulas: we will usually choose $\mu=2$. Under transformation to these new variables, the Hamiltonian can be expressed in the {\it reduced} form 
\ba
\mathcal K({\cal R}, \psi; {\cal E}) = \nu K(J_1({\cal R}, {\cal E}), J_2 ({\cal R}, {\cal E}), 2(n \theta_{1}- m \theta_{2})), 
\ea
with $\nu$ a scaling factor chosen to get the simplest expression from the quasi-canonical transformation. We obtain a family of 1-dof systems in the phase-plane ${\cal R}, \psi$, with equations of motion
\ba
{\dot {\cal R}} =& 
-\frac{\partial {\cal K}}{\partial  \psi}.\label{dr}\\
{\dot \psi} =& 
\frac{\partial {\cal K}}{\partial \cal R},\label{dpsi}
\ea
parametrized by ${\cal E}$ that is conserved because is proportional to the value of the integral of motion  (\ref{calei}).

The dynamics of the 1-dof Hamiltonian $\mathcal K({\cal R}, \psi)$ are integrable. Unfortunately, this does not necessarily implies that the solution of the equations of motion can be written explicitly. However, a quite general description of the phase-space structure of the system is possible if we know the nature of the fixed points, since these turn out to be the {\it main periodic orbits} of the unreduced system. In fact, {\it centers} (namely maxima and minima of $\mathcal K$) are associated with stable periodic orbits which parents quasi-periodic orbits with essentially the same properties, whereas {\it saddles} of $\mathcal K$ are associated with unstable periodic orbits. For the main periodic orbits of the Hamiltonian (\ref{GNF}), ${J}_{a}, {\theta}_{a}$ are true action-angle variables and so the solutions to which they correspond are known. There are two types of periodic orbits that can be easily identified by means of the fixed points of the system (\ref{dr},\ref{dpsi}):

\begin{enumerate}
      \item The {\it normal modes}, for which one of the ${J}_{\ell}$ vanishes: the solutions $
{\cal R}=\pm{\cal E}, J_{2,1}=0 $ are respectively the periodic orbit along the $x$-axis and the $y$-axis. 
         
      \item  The periodic orbits {\it in general position}, namely those solutions characterized by fixed relations between the two angles, $\psi_0 \equiv 2(n \theta_{1}- m \theta_{2})$. These are solutions of ${\dot {\cal R}}=0$ (when ${\cal R}\ne\pm{\cal E}$) and determine the corresponding solutions of
      \be\label{RS}
      {\dot \psi} = \frac{\partial {\cal K}}{\partial \cal R} \bigg\vert_{\psi_0}= 0.\ee
      For all cases treated below they fall in two classes: $\psi_0 = 0$ (to which we refer as the {\it in phase} oscillations) and $\psi_0 = \pm \pi$ (the {\it anti-phase} oscillations). 
       \end{enumerate}
As a rule, normal modes exist on every surface $K = (n/\omega_2) E$, where $E$ is the true energy. Periodic orbits in general position exist instead only beyond a certain threshold and we speak of a bifurcation ensuing from a detuned resonance. The bifurcation is usually described by a series expansion of the form

\ba\label{detexp}
 E_c = \sum_{k} c_{k} \delta^{k} = \sum_{k} c_{k} \left(q - \frac{m}{n} \right)^{k},\ea
where the $c_{k}$ are coefficients depending on the resonance ratio and the parameters of the system. Eq.(\ref{detexp}) implies that at exact resonance (vanishing detuning) the bifurcation is intrinsic in the system and that, when we move away from the exact ratio, the critical value $E_c$ of the threshold energy gradually increases.  We will see that already a linear relation given by the first order truncation provides a reliable estimate of the threshold values. Actually, by using Hamiltonian   (\ref{GNF}), the thresholds naturally appear in terms of the `distinguished' variable $\cal E$: to arrive at expressions of the form (\ref{detexp}) we need to disentangle the relation between $\cal E$ and the true energy \citep{bbp}.

By plotting curves (\ref{detexp}) in the $(q,E)$-plane we get  information for a given value of the other morphological parameters ($\alpha$ in our reference cases). Each resonance corresponds to a family of periodic orbits to which it is customary to assign the nicknames introduced by \citet{mes}. The nature of the critical points of the system (\ref{dr},\ref{dpsi}) determines the stability/instability property of the orbit. With obvious limitations due to a perturbative approach we may deduce the main aspects of the phase-space structure. We recall that the most obvious limitation of the method is determined by the values of dynamical and/or morphological parameters beyond which the dynamics are mostly chaotic. We can increase the precision in the prediction of the thresholds by adding terms to the normal form: the minimal order of truncation of the series is determined by $N_r $, that of the first resonant term in the normal form. However there is an {\it optimal} order that can be assessed by exploring the asymptotic properties of the series \citep{pbb}, but this issue is beyond the scope of the present work. In the following sections we compute the series (\ref{detexp}) in the most significative cases. 

\section{Bifurcation of the loops}\label{loop}

The system given by (\ref{Hamiltonian}) represents motion in the symmetry planes of a triaxial galaxy. In each of those planes, the symmetry axes directly give periodic orbits. Consider the models (\ref{pota}) at low energy: since the dynamics are slightly different from those of a harmonic oscillator ($\alpha = 2$), we may expect them to be stable oscillations. What happens when energy increases? Nonlinear dynamics give asynchronous motions, with frequencies depending on amplitudes. Instability can be triggered by low-order resonance and we can expect a transition to instability and the birth of new orbits. 

The most common occurrence is that of the {\it loops}, closed orbits simply encircling the origin. We are going to see that the bifurcation providing the loops can be easily described with a 1:1 resonant normal form: they correspond to the anti-phase class $\psi_0 = \pm \pi$ introduced above (there are {\it two} of them, one rotating clockwise, the other counter-clockwise). The other class of 1:1 resonant orbits, the in-phase $\psi_0 = 0$ {\it inclined} orbits, are straight segments rotated with respect to the principal axes \citep{zm} and are forbidden in the case of strictly elliptical equipotentials. Therefore we may ask ourselves how much we have to `deform' the elliptical equipotentials in order to accommodate for this class too.

\subsection{The 1:1 resonant normal form}

The general treatment of the {\it m=n}=1 symmetric resonance with two reflection symmetries has been given by \citet{zm} on the basis of previous work by \citet{fv}. Their results, based on the method of averaging, in principle contain the answer to the questions posed at the start of this section. We prefer to present these results within the framework of Lie transform normalization because it is more effective in particular when studying higher-order resonances. 

We approximate the frequency ratio with (\ref{det}) in the 1:1 case,
\ba\label{det11}
q=1+\d = 1+\tilde\d\e^2,
\ea
so that, after the scaling transformation (\ref{sca1}--\ref{sca2}) and (\ref{scah}), the Hamiltonian (\ref{Hamiltonian}) takes the form (\ref{Hr}) with
\begin{eqnarray}
H_0&=&\frac12 (p_1^2 + x_1^2 + p_2^2 + x_2^2)\\
H_2&=&\frac{\tilde\d}{2}(x_1^2+p_1^2) + B_1\left(x_1^2+x_2^2\right)^2.
\end{eqnarray}
We truncate at order $N=N_r=1$ and consistently expand the ellipticity parameter according to (\ref{dete}) up to the same order.  
A standard normalization procedure  \citep{bbp,MP} transforms the Hamiltonian (\ref{Hr}) into the `normal form'
\begin{eqnarray}
K_{11} &=& J_1+J_2+ \e^2 \tilde\d J_1 + \label{Kalpha11AA}\\
&&\e^2  B_1 \left(\frac32 (J_1^2 + J_2^2) +  J_1J_2 (2 + \cos(2 \theta_1-2 \theta_2))\right).\nonumber
\end{eqnarray}

\subsection{Bifurcation of the 1:1 resonant periodic orbits}

By introducing quasi-canonical variables adapted to the resonance by means of the linear combinations (\ref{calr}) with $\lambda=\mu=2$,
\be\label{calr11}
{\cal E}=J_1+J_2, \quad {\cal R}=J_1-J_2, \quad \psi=2 (\theta_{1}- \theta_{2}),\ee
 the normal form (\ref{Kalpha11AA}) becomes
\begin{eqnarray}
\mathcal K_{11} &=& \frac{\delta }{2}\mathcal R+\frac{B_1}4 \left(3 {\mathcal R}^2+
 ({\cal E}^2 - {\mathcal R}^2) (2 +  \cos\psi)\right),\label{KER11}
\end{eqnarray}
where constant terms have been neglected for simplicity and, since all non-constant terms are of the same order in $\e$, it too has been factored out.
 $\mathcal K_{11}$ defines a one--degree of freedom system with the following equations of motion
\ba
{\dot {\cal R}} &=&
\frac{B_1}4 (\mathcal E^2 - \mathcal R^2) \sin\psi, \label{dr1}\\
{\dot \psi} &=& \frac{\delta }{2}+\frac{B_1}{2}\mathcal R  (1-\cos\psi).\label{dpsi1}
\ea
As anticipated above, $\psi=0$ and $\psi=\pm\pi$ solve (\ref{dr1}) when ${\cal R}\ne\pm{\cal E}$. However, for $\psi=0$,  equation (\ref{dpsi1}) does not admit any solution in $\mathcal R$. This means that {\it inclined orbits do not appear}. Rather, for $\psi=\pi$, we find the solution
\ba\label{PO11L}
\mathcal R=\mathcal R_{\ell}\equiv-\frac{\delta}{2B_1}.\ea
In view of (\ref{calr11}), the constraints $ 0 \le J_1,J_2\le{\cal E}$
applied to this solution give the condition of existence for loop orbits. By using (\ref{det11}) for the ellipticity parameter, we find the threshold
\ba \label{EPOL1}
{\cal E}_{\ell}\equiv\frac{1-q}{2B_1}.
\ea
To be concrete we can express this result in the case of the $\alpha$-models (\ref{pota}). In view of the rescaling and of the expansion of the energy as a truncated series in the parameter ${\cal E}$, we have that $E = {\omega_2} {\cal E} = {\cal E} / q$ is a first order estimate of the `true' energy of the orbital motion. We can use the above critical values to establish the instability threshold for the model problem given by potentials (\ref{pota}):
\ba\label{DS11L}
E \ge E_{\ell} = \frac{4|1-q|}{2-\alpha}.\ea
In the range
\ba
0.7 < q < 1.3,\ea
which can be considered as `realistic' for elliptical galaxies, the thresholds (\ref{DS11L}) give estimates correct within a 10\% if compared to numerical computations \citep{bbp,pbb}. When \eqref{DS11L} is satisfied, loop orbits bifurcate from the $y$-axial normal mode in the oblate case, and from the $x$-axial normal mode in the prolate case \citep{MP}. At the same bifurcation values, the normal mode suffers a change of stability, passing from stable to unstable when the new orbit is born. By direct check of the nature of the critical point $(\mathcal R=\mathcal R_{\ell},\psi=\pi)$ of the function (\ref{KER11}) the loop, when it exist, is stable.

To get a higher precision, we have to include higher-order terms in the series expansion.
If we expand the potential up to order six and truncate the normal form at $N=2$, the critical energy \eqref{DS11L} up to order two in the detuning parameter is given by
\begin{eqnarray}
E_{1\ell}&=&\frac{4  }{2 - \alpha}(1-q)+\frac{2 (2+3 \alpha )}{(\alpha-2 )^2} (1-q)^2  \\
E_{2\ell}&=&-\frac{4 }{2 - \alpha} (1-q)+\frac{2 (5 \alpha -2)}{(\alpha-2 )^2} (1-q)^2
\end{eqnarray}
respectively in the oblate and prolate case. These generalize the expression for the logarithmic ($\alpha=0$) potential reported in \citep{bbp}. We don't give the details to arrive at these results since the second order case is explicitly treated in Section \ref{ban} where it is necessary to describe the 1:2 resonance.

\subsection{Ellipse-breaking deformations}\label{ebd} 

Let us now consider a deformation of the potentials \eqref{pota}  by introducing a small parameter $\beta$ such that
\begin{eqnarray}\label{ebdp}
\mathcal V_{2}&=&B_1 (s^4+2\beta x^2 y^2),
\end{eqnarray}
with `boxy' or `disky' shapes of the level curves when respectively $\beta < 0$ and  $\beta > 0$. 
As we will show in the following, the presence of the parameter $\beta$ affects the bifurcation of
inclined orbits. The normal form of the system is the same as $ {\cal K}_{11}$ in (\ref{KER11}) except that the coefficient in front of the resonant term is replaced by
\be
B_1 \left(1 + \beta \right).\ee
The important point is that the second equation of motion for the reduced system becomes
\begin{eqnarray}
{\dot \psi} &=& \frac{\delta }{2}+\frac{B_1}{2}\mathcal R \left(3 - (1 + \beta)(2+\cos\psi)\right).\label{dpsib} 
\end{eqnarray}
Now, for $\psi=0$  equation \eqref{dpsib} admits the solution 
\ba
\mathcal R=\mathcal R_i(\beta)\equiv\frac{\d}{3B_1\beta}.
\ea
This fixed point determines two inclined orbits for the original system.  For $\psi=\pi$ the right-hand side of equation  \eqref{dpsib} vanishes for
\ba
\mathcal R=\mathcal R_{\ell}(\beta)\equiv-\frac{\d}{B_1(2-\beta)}.
\ea
This determines the loops as before and is only slightly changed with respect to the solution (\ref{PO11L}) found above. 

Working as usual in the family  \eqref{pota}, the constraints $ 0 \le J_1,J_2\le{\cal E}$
translate into the existence condition
\ba \label{enib}
\mathcal E \geq\mathcal E_{1,2i}(\beta)\doteq\pm\frac{4(1-q)}{3(2 - \alpha)\beta}.
\ea
and
\ba\label{enlb}
\mathcal E\geq\mathcal E_{1,2\ell}(\beta)\doteq\pm\frac{4(1-q)}{(2 - \alpha)(\beta-2)},
\ea
where now, with the indexes $1,2$, we now distinguish between the bifurcations from the two normal modes. The critical values \eqref{enib} correspond respectively to the bifurcation of  inclined  orbits from the $y,x$-axial normal mode and the same with \eqref{enlb} for the loops. This distinction is relevant if one is interested in which normal mode suffers a change of stability when a new orbit arises. 

Thus, if we break the ellipticity of the potential, {\it inclined orbits appear}: however the smaller the deformation, the higher the threshold value \eqref{enib}. Loops continue to bifurcate at a lower energy: to change the bifurcation sequence, unreasonable high values of $\beta$ are required. The phenomenon is anyway interesting because it can easily be checked that the two families are always of different stability nature: the stable one is the first to appear, therefore there is a critical value of $\beta$ at which there is an exchange of stability between loops and inclined. The special value $\beta=2$ producing the singularity in  \eqref{enlb} is associated to exact separability in rotated Cartesian coordinates which forbids the existence of the loops.

One may wonder if the inclusion of additional terms in the series does modify qualitatively the results obtained at lower orders: a nice result provided by the theory of singularity \citep{BLV} proves that this is not the case, at least for the symmetric 1:1 resonance. The case with elliptical equipotential ($\beta=0)$  is in a certain sense {\it degenerate}, but a generic symmetry-preserving deformation is stable. The meaningful information is essentially contained in the normal form truncated at $N=1$ since, even adding higher-order terms to the original physical Hamiltonian, one can always find a non-linear coordinate transformation allowing us to eliminate the extra terms from the normal form: in other words the bifurcations predicted by using (\ref{KER11}) (including the deformation) are qualitatively reliable and can only be quantitatively improved with a higher-order normalization \citep{pbb}. 

\section{Bifurcation of the banana and anti-banana}\label{ban}

Another important class of bifurcations is that of {\it banana} orbits \citep{mes} usually associated to the instability of the major-axis orbit. It corresponds to a pair of in-phase ($\psi_0 = 0$) oscillations with frequency ratio 1:2. The anti-phase family are the figure-eight periodic orbits, or {\it anti-banana}: we will show that in the potentials  \eqref{pota} {\it stable} bananas bifurcates at lower energies than {\it unstable} anti-bananas for relevant values of the parameters.

%\subsection{1:2 Symmetric Resonance}
In the case of the  {\it m}=1,{\it n}=2 resonance with reflection symmetries about both axes,
we know from the general expression \eqref{GNF} of the normal form, that the normalization procedure must be pushed at least to order $N_r=2$.
The terms in the series expansion \eqref{Hr} are now given by
\ba
H_0&=& \frac12 m(p_1^2 + x_1^2)+p_2^2 + x_2^2,\\
H_2&=&\tilde\d(x_1^2+p_1^2) + B_1 (x_1^2 + 2 x_2^2)^2, \\
H_4&=& 2\tilde\d B_1 (x_1^4 - 4 x_2^4) + B_2 \left(x_1^2 + 2 x_2^2\right)^3.
\ea
After normalization, we get the `normal form'
\be \label{12nf}
K_{12}=\sum_{k=0}^2 \e^{2k}K_{2k}, 
\ee
where
\ba
K_0&=&J_1+2J_2, \\
K_2&=&2\tilde\d J_1 + B_1 \left( \frac32 J_1^2 + 4 J_1 J_2 + 6 J_2^2 \right), \\
K_4&=& 3 \tilde\d B_1 \left( J_1^2  - 4 J_2^2 \right) - (17 B_1^2 - 10 B_2) \left( \frac14 J_1^3  +2 J_2^3 \right) \nn
& -&
 \frac23 (46 B_1^2 - 27 B_2) J_1 J_2^2  - \left(\frac{56}3 B_1^2 - 9 B_2 \right) J_1^2 J_2 \nn
& -&
    \frac32 (2 B_1^2 - B_2) J_1^2 J_2 \cos(4 \theta_1 - 2 \theta_2).
\ea
We remark that in the computation of \eqref{12nf} and results thereof, the use of algebraic manipulators like Mathematica\textregistered \
is practically indispensable. 

The quasi-canonical transformation to adapted resonance coordinates now is
\be\label{calr12}
 \left\{\begin{array}{ll}
  J_1=&\E+2\R  \\
   J_2=& 2\E-\R \\
\psi=&4\theta_{1}- 2\theta_{2} \\
\chi=&2\theta_{1}+ 4\theta_{2}
\end{array}\right.\ee
and the effective Hamiltonian 
\be
\mathcal K_{12}(\mathcal R, \psi; \E) \doteq K_{12} (J_a(\E,\R),\theta_{a}(\psi,\chi))\ee
defines the following equations of motion
\ba
\dot\R &=& -\frac32 \e^4 (2 B_1^2 - B_2) (2\E - \R) (\E + 2 \R)^2 \sin\psi,     \label{rpunto12}\\
\dot\psi&=& 2 \e^2 \left(B_1(3 \E - 4 \R) - 2 \tilde\d \right) + \nn
&-&\frac16 \e^4 \left[5 {\cal A}(\R;\E,\tilde\d) - {\cal B}(\R;\E) \cos\psi \right],
\label{psipunto12}
\ea
 where
 \ba
 {\cal A} &=&  36 B_2 \E (-3 \E + 4 \R)\nn 
 &+& B_1^2 (155 \E^2 - 276 \E \R + 48 \R^2) + 
   72 B_1 \E \tilde\d,\nn
  {\cal B} &=& 9 (2 B_1^2 - B_2) (7 \E^2 + 8 \E \R - 12 \R^2).\nonumber\ea
  
 The fixed points of this system give the periodic orbits of the original system. The pair of solutions with ${\cal R}=2\E$, $\R=-{\E}/{2}$ respectively correspond to the normal modes along the $x$-axis  and $y$-axis. Let us look for periodic orbits in general position. We start with setting $\psi=0$ and looking for $\R$-solutions of
$\dot\psi=0$. Since we are dealing with a perturbation problem  in $\e$, we look for a solution in the form  \citep{henrard}
\ba\label{Rtb}
\R=\R_0+\R_1 \e^2+O(\e^4).
\ea
We substitute \eqref{Rtb} in  \eqref{psipunto12} with $\psi=0$ and collect terms up to fourth order in $\e$. Equating to zero the coefficient of second order, we find that $\R_0$ has to satisfy
\ba
B_1(3 \E - 4 \R) - 2 \tilde\d=0
\ea
which gives
\ba
\R_0 \equiv \frac34 \E - \frac{\tilde\d}{2 B_1}.
\ea
Once computed $\R_0$ we find the coefficient of the second order term in the expansion of the fixed point
\ba\label{cpb}
\R_{b}\equiv\R_0+\R_{b1}\e^2,\;\;\;\psi=0,
\ea
which determines the banana orbits:
\ba
J_{1b}=\E+2\R_{b}, \\
J_{2b}=2\E-\R_{b}.
\ea
Similarly, for $\psi = 4\theta_{1}- 2\theta_{2} = \pi$, we find a solution of the form 
\ba\label{cpa}
\R_{a}\equiv\R_0+\R_{a1}\e^2,\;\;\;\psi=\pi,\ea 
and $J_{1a}=\E+2\R_{a}, J_{2a}=2\E-\R_{a},$
corresponding to the antibanana orbits.

In view of (\ref{calr12}), the constraints $ 0 \le J_1\leq5\E$, $0\leq J_2\leq{5 {\cal E}}/{2}$
applied to these solutions give the condition of existence for these periodic orbits. Non trivial existence conditions can be found by solving $J_{1,2b}\geq0$ for the bananas and $J_{1,2a}\geq0$ for the anti-bananas. The implicit function theorem
assures that there exists unique solutions $\E_c=\E(\d)$ in each cases determining the bifurcation thresholds. For the bananas, up to the second perturbative order we get
\ba\label{EPOB}
{\cal E}_{b1}&=&-\frac{2}{5 B_1}\tilde\d + \frac{59 B_1^2 - 27 B_2}{15 B_1^3} \tilde\d^2 \e^2,\label{EPOB1}\\
{\cal E}_{b2}&=&\frac{2}{5 B_1}\tilde\d + \frac{97 B_1^2 - 36 B_2}{15 B_1^3} \tilde\d^2 \e^2,\label{EPOB2}
\ea
which respectively determine the bifurcation from the $x$-axial normal mode in the first case, and from the $y$-axial normal mode in the second case (we discuss below which of these possibilities actually shows up). Similarly, the threshold values that
gives the existence condition of anti-banana orbits are given by
\ba
{\cal E}_{a1}&=&-\frac{2}{5 B_1}\tilde\d + \frac{19 B_1^2 - 9 B_2}{3 B_1^3} \tilde\d^2 \e^2,\label{EPOA1}\\
{\cal E}_{a2}&=&\frac{2}{5 B_1}\tilde\d + \frac{97 B_1^2 - 36 B_2}{15 B_1^3} \tilde\d^2 \e^2.\label{EPOA2}
\ea
By comparing \eqref{EPOB2} with \eqref{EPOA2} we see a first interesting result: if the bifurcation occur from the $y$-axis, {\it banana and anti-banana appear together}. It is therefore important to discriminate between the two possibilities. Since the dominant term in the series is the first and $\E$ must be positive, we see that case 1 (bifurcation from the $x$-axis) or 2  (bifurcation from the $y$-axis) occur if $\tilde\d$ and $B_1$ have different sign or not. To write the expressions of the bifurcation curves in the physical $(q,E)$-plane, according to the rescaling \eqref{scah} with $n=2$, on the two axial orbits we have
\ba
E_1&=&5 \E \e^2 + \frac{75}{2}B_1 \E^2 \e^4+ O(\e^6),\\
E_2&=&5 \E \e^2 + \left(\frac{75}{2}B_1 \E^2 - 10 \E \tilde\d \right) \e^4+ O(\e^6),
\ea
so that we get 
\ba
E_{b1}&=&-\frac{2}{B_1}\d + \frac{77 B_1^2 - 27 B_2}{3 B_1^3} \d^2,\\
E_{a1}&=&-\frac{2}{B_1}\d + \frac{113 B_1^2 - 45 B_2}{3 B_1^3} \d^2,
\ea
for the bifurcations from the $x$-axis and
\ba
E_{b2}&=&E_{a2}=\frac{2}{B_1}\d + \frac{103 B_1^2 - 36 B_2}{3 B_1^3} \d^2,
\ea
for the bifurcations from the $y$-axis. To be concrete, for our family \eqref{pota} we have that, with $\alpha > 0$, the coefficient $B_1$ is negative. The ellipticity  is usually $q>1/2$ so that $\d>0$, therefore relevant thresholds are
\ba
E_{b1}&=& \frac{16}{2-\alpha }\left( q - \frac12 \right)+\frac{8 (41 \alpha-10 )}{3 (2-\alpha )^2}\left( q - \frac12 \right)^2, \label{E2BA} \\
E_{a1}&=&\frac{16 }{2-\alpha }\left( q - \frac12 \right)+\frac{8 (53 \alpha+14 )}{3 (2-\alpha )^2}\left( q - \frac12 \right)^2. \label{E2AA}
\ea
Since the difference 
\be
E_{a1}-E_{b1}=32\frac{2+\alpha}{(2-\alpha )^2}\left( q - \frac12 \right)^2
\ee
is positive, we verify that, for models in the class  \eqref{pota} and with parameter ranges useful for elliptical galaxies, the bifurcation sequence is always from the major axis, with bananas appearing at lower energies than anti-bananas. By checking the nature of the two critical points (\ref{cpb},\ref{cpa}), it can be seen (it is a tedious but straightforward computation,  \citet{MP12}) that in systems \eqref{pota} the first family is always stable and the second unstable: (\ref{E2BA},\ref{E2AA}) generalize the corresponding expressions for the logarithmic ($\alpha=0$) potential reported in \citet{bbp}. As long as the banana does not bifurcate the major axis is stable and parents `box' orbits. It loses its stability at the first bifurcation and regains it at the second. It  is natural to ask how much these results are affected by ellipse-breaking deformations: we can say that, in analogy with what seen for the 1:1 resonance, the hierarchy of bifurcations changes only for unreasonable high values of the deformation parameter.

\section{Higher-order symmetric resonances}

As is well known \citep{mes} stable periodic orbits corresponding to higher-order resonances and quasi-periodic orbits parented by them give a small but not-negligible contribution to regular dynamics in systems with cores. In realistic cases with mixed (regular+chaotic) dynamics it is conjectured that these `boxlets' may become important in shaping the bulk of the density distribution \citep{Zhao,ZZ}. The main difference of these families from those seen above consists of the fact that their bifurcation is not connected with the loss (or regain in case of a second bifurcation) of stability of the normal mode. The birth of periodic orbits with $N_r > 2$ is rather due to breaking of a resonant torus around the normal mode and is correctly described by applying the Poincar\'e-Birkhoff theorem  \citep{Arn}: however, the technique we applied above continues to work and the conditions for the existence and stability of an $m/n$-resonant periodic orbit with $m+n>3$ can still be found by constructing the appropriate normal form and locating fixed points of the reduced system. 

A technical issue worth to be clarified is the following: by reducing the resonant normal form \eqref{GNF} truncated at order $N_r$ by means of the transformation \eqref{calr}, we obtain a polynomial of degree $N_r+1$ in $\R$. The corresponding equation of motion for $\psi$ produces a pair of algebraic equations of degree $N_r$ which have to be solved to locate the fixed points (one for each solutions $\psi_0$, cfr. point ii in Sect.2). This problem is very difficult to solve if, for $N_r > 2$, we aim at general solutions depending on the parameters of the system. However we are not interested in every solution but only in those connected with the passage through the chosen resonance. We can therefore resort to the perturbation method we have described in detail in the previous section on the 1:2 resonance. In that case, with $N_r=2$ we had to solve two equations of second degree (cfr. the rhs side of \eqref{psipunto12}): this clearly does not represent a problem since we can write explicitly the two pair of solutions. However, in each pair, {\it only one} solution is geometrically acceptable because it satisfies the condition at resonance; the other must be discarded by direct check. The perturbative method based on the construction of the series \eqref{Rtb} \citep{henrard} automatically selects the acceptable solution. The method is therefore extremely useful for higher-order resonances: a solution of the form
\ba\label{Rtbn}
\R=\sum_{k=0}^{N_r-1} \R_k \e^{2k}+O(\e^{2 N_r})
\ea
easily allows us to select the meaningful solution without any loss in accuracy.

We have applied the method to the case of {\it fish} orbits corresponding to the (anti-phase) 2:3 resonance. In this case, $N_r = 4$: the  Hamiltonian series must be expanded up to include terms of degree 10 ($B_4$ in the original potential). The explicit expressions of the normal form in the general class  \eqref{seriese} and for the family \eqref{pota} are a bit heavy to write and are reported elsewhere \citep{AM12}: they are available upon request as Mathematica\textregistered \
notebooks. Anyway the procedure is a straightforward extension of that illustrated in the previous section.

The threshold for the existence of fish orbits turns out to be
\ba
E_{f}&=&-\frac{3}{2 B_1}\d + \frac{9}{80 B_1^3} (149 B_1^2 - 60 B_2)\d^2\nn
&-&\frac{27 (7671 B_1^4 - 7840 B_1^2 B_2 + 
 3600 B_2^2 - 1500 B_1 B_3) \d^3}{1600 B_1^5}\nn
 &+&\frac{81}{448000 B_1^7} (4852431 B_1^6 - 8889450 B_1^4 B_2  \nn
 &&+ 
 9116400 B_1^2 B_2^2-3780000 B_2^3 - 3626000 B_1^3 B_3 \nn
 && + 
 3150000 B_1 B_2 B_3 - 490000 B_1^2 B_4)\d^4.
\ea
This result is undoubtedly unpleasant to write (and read!) but it testifies what is the rule with high-order expansions. However, trusting the normalization program and paying attention to write down the results without errors, the series give us numbers we can use in specific cases. In terms of the parameters of the family \eqref{pota} we get
\ba
E_{f}&=&\frac{12}{2 - \alpha}\d - \frac{9 (22 + 69 \alpha)}{10 (2 - \alpha)^2} \d^2\nn
&+&\frac{9 (4372 + 2508 \alpha + 4853 \alpha^2) \d^3}{200 (2 - \alpha)^3}\nn
 &+&\frac{27(1368856 + 3109116 \alpha + 542642 \alpha^2 +1468293 \alpha^3)\d^4}{56000 (2 - \alpha)^4} \nn
\label{E23}\ea
where in this case
\ba
\d=q-\frac23.\ea
This result completes and generalizes the treatment of the logarithmic case presented in \citet{bbp}. We may ask if it is worth the effort: in the logarithmic case ($\alpha=0$), \citet{mes} numerically found $E_f (q=0.7) = 0.21$ and $E_f (q=0.9) = 2.28$ that we can consider {\it experimental} exact threshold values. Our analytic result predicts $E_f (q=0.7) = 0.206$ and $E_f (q=0.9) = 2.10$. The agreement is excellent near the resonance ($\d=0.7-2/3\simeq0.03$) and only moderate further away from it ($\d=0.9-2/3\simeq0.23$). However, we remark that the energy level $E=2.28$ is extremely high if seen with the eye of the perturbation theorist: an error of 8\% may then appear not so bad. Moreover it is possible to improve the quality of the prediction by going to still higher orders.  

If one is only interested in a rough prediction around a general {\it m}:{\it n} resonance \citep{P09}, from these results we can deduce the general first order expression 
\ba
E_{m:n}&=&\frac{n}{m B_1}\d ,\ea
that, for the family \eqref{pota}, gives
\ba\label{EMN}
E_{m:n}&=&\frac{8n}{m (2-\alpha)} \left(q-\frac mn \right).\ea
The example of the 3:4 resonance (the {\it pretzel}) is a good test: for the logarithmic potential \citet{mes} numerically found $E_{3:4} (q=0.7) = 0.25$ and $E_{3:4} (q=0.9) = 1.22$. Eq.\eqref{EMN} with $\alpha=0$ predicts $E_{3:4} (q=0.7) = 0.27$ and $E_{3:4} (q=0.9) = 0.80$. The agreement is quite good near the resonance; moreover, since for $q < 3/4$ \eqref{EMN} is negative, accordingly with the treatment of the previous cases, we may predict that in the case $q=0.7$ the `bifurcation' is from the $y$-axis, as actually found by \citet{mes}. 

\section{Discussion}

In the present section we discuss some implications of the results described above and present open problems and possible directions to cope with them. Here we also recall that the approach we have followed is not the only possible and that both the normalization and the reduction can be obtained by exploiting alternatives like the {\it Lissajous transformation} \citep{DE}, the method of geometric invariants \citep{HSS} and the singularity theory \citep{BLV} mentioned in the introduction.

\subsection{Asymptotic expansions}\label{asy}

Series like those described in this work are asymptotic: this means that a truncation of the series, say at order $N$, apparently converges in a given domain only for $N<N_{\rm opt}$, the {\it optimal} truncation order linked to the extent of the domain. We remark that this semi-convergence is in general not associated to a true function: rather, it is only associated to a local geometric object we use as an invariant surface in the regular part of phase space. The optimal truncation depends on the problem at hand and to assess it a priori is quite difficult \citep{ECG}. In \citet{pbb} we have tried to estimate $N_{\rm opt}$ for two members of the family \eqref{pota}, $\alpha=0,1$. For the bifurcation of the banana in the logarithmic potential, we obtained $N_{\rm opt} > 7$ for $q\le0.7$,  $N_{\rm opt} = 6$ for $q=0.8$ and $N_{\rm opt} = 3$ for $q=0.9$. In this case (the worst being the furthest from exact resonance) the relative error of the prediction is 11\%. However, the quality of the prediction (and the corresponding optimal order) can be further improved if different techniques of summation are employed. \citet{scu}  suggested to use the continued fraction method \citep{b3} to re-sum asymptotic series: we applied this idea to the bifurcation threshold series and in  the worst case just mentioned (banana with $\alpha=0,q=0.9$) we got $N_{\rm opt} = 5$ lowering the relative error to less than 4\%. What is indeed remarkable in this result is that the bifurcation energy is $E=3.6$. For the logarithmic potential this corresponds to a radius of order 40 times larger than the convergence radius of the original series \eqref{seriese} so that we have an outstanding evidence of the power of asymptotic expansions. 

\subsection{St\"ackel fits}

Separable systems play an important role among integrable systems since they provide explicit solutions for the orbit structure. The application of St\"ackel systems to approximate the dynamics of galaxies is therefore a classical field \citep{VH,dzb,kent,VZB}. \citet{ZL} proposed a `St\"ackel fit' of galactic potentials around an equilibrium to take advantage of the opportunity of exploiting the integrals of motion of systems separable in elliptical coordinates. At order $N=1$ the number of free parameters is sufficient to fit any expansion; at higher orders the fit is constrained by conditions on the coefficients. The method works since the dynamics of a St\"ackel system separable in elliptical coordinates resemble that of the 1:1 resonance for potentials of the form \eqref{seriese}: however, the results obtained in Subsection \ref{ebd} warn us from excessive confidence in the method. In fact we can fit a potential of the form \eqref{ebdp} or even more general in which inclined orbits may play a role: however the fitting St\"ackel potential {\it does not support inclined}. Although this may not be of particular relevance in galactic applications, it is a problem as a matter of principle. We remark that St\"ackel systems do not end with those mentioned above, but include those separable in other coordinate systems. In 2 dimensions, separability in parabolic coordinates can be used to model elliptical disks \citep{ST1,ST2}: in this case there is a relation with the 1:2 resonance. However, systems separable in parabolic coordinates accommodate bananas and quasi periodic orbits parented by them, but do not support their anti-phase companions.

\subsection{Surfaces of section}

By inverting the transformation leading to the normal form we can compute {\it formal} integrals of motion \citep{contos,conto2} which have to be interpreted as asymptotic series as prescribed in Subsection \ref{asy}. The most immediate use of these expansions is to construct approximations of Poincar\'e surfaces of sections: for the logarithmic potential, \citet{bbp} show that, at sufficiently high energy, surfaces constructed around low-order resonances display a quite close resemblance with those numerically obtained in the scale-free limit by \citet{mes}. Moreover, by using asymptotic series as true phase-space conserved functions in a suitable domain, bifurcation curves can be computed by investigating the nature of the critical points of these functions. The results are identical to those obtained with the normal form when expressed as series in the detuning: either approaches being effective, one can chose which minimize the computational effort.

\subsection{Order and chaos}

The domain of `semi-convergence' of asymptotic series approximating invariant surfaces of generic systems can be taken as a measure of their regular dynamics. We have seen that, as a matter of principle, regular phase-space zones associated to resonances of any order can be adequately included and described. The approach to high-order resonances is dual: either their role is considered to be marginal \citep{sa} or they are considered as an inescapable signature of chaos \citep{bt}. However, in several interesting cases (see e.g. the scale-free models with $\alpha>0$ treated by \citet{TT}) we have that different resonances coexist without overlapping for a large range of parameters. Resonance manifolds generate a structure that can be understood via reduction \citep{tv}. 
Regular dynamics are `complicated' but definitely not chaotic, so efficient tools to investigate their features are extremely useful.

\subsection{3D models}

The most relevant generalization is towards 3 dimensional systems. The pioneering work by \citet{dza} still remains a major contribution since mathematicians, although have devoted much effort to this issue, analyzed in general only simple abstract models \citep{SV}. \citet{dza} gave an almost complete study of the orbit structure of a generic quartic potential around the 1:1:1 resonance. The relevance of this case is testified by the fact that, in spite of a radical change in our understanding of elliptical galaxies with cusps affecting their overall dynamics, the two orbit families characterizing triaxial systems are still considered to be the boxes and the long axis-tubes \citep{vbz}: we therefore see that the study of the stability of the $x$-normal mode and the condition for existence of stable loops in the $yz$-plane as studied in this work is very useful.

The main problem with 3 degrees of freedom is that the normal form itself is in general not integrable: the normalization procedure of resonant Hamiltonians provides only {\it one} formal integral \citep{gu} in addition to energy. However, the study of the stability of the three normal modes and the bifurcations of periodic orbits in general position can be done even in the absence of a third integral. The step towards a general analysis of relevant cases like the 1:2:2 and 1:2:3 resonances seems to be within the reach of the method. We also recall that a small bulk rotation of the ellipsoid can be included with a suitable canonical transformation \citep{zm}.

\section{Conclusions}

We have presented a general analysis of the orbit structure of 2D potentials with self-similar elliptical equipotentials. The main results are the following:

The 1:1 resonance is associated to the appearance of the loops and leads to the destabilization of the $y$-axis orbit in the oblate case and of the $x$-axis orbit in the prolate case. Inclined orbits are never present and may appear only when the equipotentials are heavily deformed.

The 1:2 resonance determines the appearance of bananas and anti-banana orbits: the first family is stable and always appears at a lower energy than the second, which is unstable. The bifurcation sequence produces the change in the stability character of the major axis orbit and is modified only by very large deformations of the equipotentials.

Higher-order resonances appear at intermediate energies which can be predicted with good accuracy.

We have analyzed several issues connected with the approach and sketched the directions for further work. In particular, we think that evaluating the overall predictive power of the method based on asymptotic expansions is a decisive step if one is interested in studying stationary or rotating triaxial potentials.

\label{lastpage}

\end{document}